\documentclass[pra,twocolumn,showpacs]{revtex4-1}

\usepackage{graphicx}
\usepackage{amsmath}
\usepackage{color}
\usepackage{float}

\newcommand{\comment}[1]{}

\newcommand{\tr}[2][]{\text{Tr}_{#1}\left\{#2\right\}}

\begin{document}

\title{Time evolution of open quantum many-body systems}

\author{Vincent R. Overbeck}
\email[]{vincent.overbeck@itp.uni-hannover.de}
\author{Hendrik Weimer}

\affiliation{Institut f\"ur Theoretische Physik, Leibniz Universit\"at Hannover,
  Appelstra{\ss}e 2, 30167 Hannover, Germany}

\begin{abstract}

  We establish a generic method to analyze the time evolution of open
  quantum many-body systems. Our approach is based on a variational
  integration of the quantum master equation describing the dynamics
  and naturally connects to a variational principle for its
  nonequilibrium steady state. We successfully apply our variational
  method to study dissipative Rydberg gases, finding very good
  quantitative agreement with small-scale simulations of the full
  quantum master equation. We observe that correlations related to
  non-Markovian behavior play a significant role during the relaxation
  dynamics towards the steady state. We further quantify this
  non-Markovianity and find it to be closely connected to an
  information-theoretical measure of quantum and classical
  correlations.

\end{abstract}


\maketitle

\section{Introduction}
Understanding the time evolution of quantum many-body systems is
currently one of the most challenging tasks in both atomic and
condensed matter physics, as concepts and tools developed for
equilibrium systems are largely inapplicable. Open quantum systems
exhibit even greater inherent complexity owing to the necessity to
describe these systems in terms of statistical
ensembles~\cite{Breuer2002}. Here, we present a completely generic
approach to analyze the time evolution of open quantum many-body
systems, based on a variational principle.

The tremendous experimental progress in the ma\-ni\-pu\-la\-tion of
atomic quantum gases has allowed physicists to reach a regime where strong
interactions can be combined with controlled dissipative processes
\cite{Syassen2008,Baumann2010,Barreiro2011,Krauter2011,Barontini2013},
offering the potential to dissipatively prepare novel classes of
quantum many-body states
\cite{Diehl2008,Verstraete2009,Weimer2010}. Recently, this
experimental progress has been especially pronounced in the context of
strongly interacting Rydberg gases
\cite{Raitzsch2009,Carr2013,Malossi2014,Schempp2014,Urvoy2015,Weber2015},
as the interaction and dissipation properties of Rydberg atoms can be
widely tuned \cite{Low2012}. Consequently, dissipative Rydberg gases
have emerged as an ideal environment to study strong interactions in
an open quantum many-body system
\cite{Lee2011,Honer2011,Glatzle2012,Ates2012a,Lemeshko2013a,Hu2013,Honing2013,Otterbach2014,Sanders2014,Hoening2014,Marcuzzi2014}.

In this paper, we present a generic framework to investigate
the time evolution of open quantum many-body systems. Our treatment
naturally connects to a variational principle for the nonequilibrium
steady state of the dynamics \cite{Weimer2015}, and allows us to reduce
the exponentially diverging number of degrees of freedom of the full
time evolution to a small number of relevant parameters. We apply our
method to the both experimentally and theoretically important problem
of dissipative Rydberg gases and compare our results to small scale
numerical simulations of the full quantum many-body problem. In our
analysis, we find that the dynamics is inherently linked to the
appearance of non-Markovian behavior within these systems, which
persists even in the nonequilibrium steady state of the
evolution. Remarkably, we observe that this non-Markovianity is
closely related to an information-theoretical measure of the quantum
and classical correlations present in the system.
\section{Variational treatment of a Rydberg gas}
We consider the dynamics of open quantum systems, which is described
in terms of a quantum master equation for the density operator $\rho$,
according to a first order differential equation $\frac{d}{dt} \rho =
\mathcal{L}\rho$, with the Liouvillian $\mathcal{L}$ being the
generator of the dynamics. To be explicit, we focus on the case where
the Liouvillian is given in Lindblad form, i.e.,
\begin{align}
 \mathcal{L}\rho=-i[H,\rho]+ \sum\limits_{i} \left(c_i\rho c_i^{\dagger}-\frac{1}{2}\{c_i^\dagger c_i,\rho\} \right),
\end{align}
with the Hamiltonian operator $H$. The
dissipative terms in the Liouvillian are given in terms of jump
operators of the form $c_i = \sqrt{\gamma} \sigma_-^{(i)}$, describing
the spontaneous decay of the Rydberg state with a decay rate
$\gamma$.
In the following, we will be interested in a numerical integration of
the quantum master equation. Here, we will employ the implicit
midpoint method \cite{Suli2003} with time $t$ and integration step size $\tau$
\begin{align}
  \rho(t+\tau) = \rho(t) + \frac{\tau}{2}\mathcal{L}\left[\rho(t) + \rho(t+\tau)\right] + O(\tau^3),
\end{align}
which we find to give the best tradeoff between accuracy of the
integration and computational cost. Within our variational approach,
we parametrize the density operator $\rho(t+\tau)$ by a set of
variational parameters $\{\alpha_i\}$. Then, we find the variational
solution by minimizing the variational norm $D$ given by
\begin{equation}
  D \equiv ||\rho(t+\tau)-\rho(t)-\frac{\tau}{2} \mathcal{L}\left[\rho(t) + \rho(t+\tau)\right]||_1 \rightarrow \min,
\label{D}
\end{equation}
where $||\cdot||_1$ denotes the trace norm given by
$\text{Tr}\{|\cdot|\}$. Crucially, the state reached in the long time
limit satisfying $\rho(t+\tau) \to \rho(t)$ corresponds to a direct
variational principle for the nonequilibrium steady state
\cite{Weimer2015}. We also note that our variational treatment is
similar to recent approaches to treat the time evolution of
one-dimensional systems based on matrix product states
\cite{Transchel2014,Cui2015,Mascarenhas2015}. In higher dimensions,
previous approaches to analyze the time evolution of open quantum
many-body systems have largely been restricted to a mean-field
decoupling \cite{Tomadin2010,Poletti2013,Vidanovic2014};
however such a mean-field treatment is problematic
  for open systems
  \cite{Hoening2014,Weimer2015,Weimer2015a,Maghrebi2015,Mendoza-Arenas2015}.


In general, the computation of the variational solution is still an
exponentially complex task. However, it is possible to obtain upper
bounds to the variational solution, which can be calculated
efficiently and are known to produce reliable results for the steady
state \cite{Weimer2015,Weimer2015a}. As a basic example, let us
consider a system consisting of $N$ identical two-level systems,
interacting by a nearest-neighbor interaction. If we restrict our
variational manifold to the set of product states, we may write the
density operator as
\begin{equation}
  \rho^p = \prod\limits_{i=1}^N \left(\frac{1}{2} + \sum\limits_{\mu\in \{x,y,z\}} \alpha_\mu^{\phantom \dagger} \sigma_\mu^{(i)}\right),
  \label{varstate}
\end{equation}
where the $\sigma_\mu^{(i)}$ are the Pauli matrices. Then, we find an upper bound to the variational norm to be \cite{Weimer2015,Weimer2015a}
\begin{equation}
  D \leq \sum_{\langle ij \rangle}||\rho_{ij}(t+\tau)-\rho_{ij}(t)-\tau \mathcal{L}\left[\rho_{ij}(t) + \rho_{ij}(t+\tau)\right]||_1,
  \label{eq:dfunc}
\end{equation}
where $\rho_{ij} = \tr[\not{i}\not{j}]{\rho}$ is the reduced two-site
density operator of the particles $i$ and $j$. 
Variational states including nearest-neighbor correlations can be expressed as
\begin{equation}
  \rho^c = \prod\limits_{i=1}^N \rho_i+\sum_{\langle ij \rangle}  \mathcal{R} C_{ij}+\sum_{\langle ij \rangle \ne \langle kl \rangle} \mathcal{R} C_{ij}C_{kl}+ \cdot \cdot \cdot 
\end{equation}
with the super-operator $ \mathcal{R}$ transforming $1_i$ in $\rho_i$
and nearest-neighbor correlations $C_{ij}$ defined as $\rho_{ij} =
\rho_i \tensor \rho_j + C_{ij}$. We can express the reduced density operator $\rho_{ij}$ in terms of the variational parameters as
\begin{equation}
\rho_{ij}^c= \frac{1}{4} + \sum\limits_{\mu, \nu} \alpha_{\mu \nu}^{\phantom \dagger} \sigma_\mu^{(i)} \otimes \sigma_\nu^{(j)}.
  \end{equation}
In the case of such correlated variational states, the corresponding upper bound to the variational norm is found to be 
\begin{equation}
 D \leq \sum\limits_{\langle ijk \rangle}||\rho_{ijk}(t+\tau)-\rho_{ijk}(t)-\tau \mathcal{L}[\rho_{ijk}(t) + \rho_{ijk}(t+\tau)]||_1,
  \label{eq:dfuncc}
\end{equation}
where $\rho_{ijk} = \tr[\not{i}\not{j}\not{k}]{\rho}$ is the reduced
three-site density operator. In a translationally invariant system,
the minimization reduces to a single three-site problem and can be
solved with only a small number of variational parameters, even for
infinitely large systems. The right hand side of Eq.~(\ref{eq:dfuncc})
is a measure of the error made during the integration arising from the
restriction of the dynamics to the variational manifold. In our
analysis, this error is always the most important one, e.g., errors
from the finite step size $\tau$ are negligible in comparison.

In the following, we will exemplify
the application of this variational principle for the time evolution
for open quantum systems, by investigating dissipative Rydberg
gases. To be specific, we consider a two-dimensional spin $1/2$ model,
with the electronic ground state and a single Rydberg state
corresponding to the two spin states. We neglect van der Waals
interactions between the Rydberg states beyond the nearest-neighbor
distance. Then, we can express the Hamiltonian of the system as
\begin{align}
  H=\frac{\Omega}{2}\sum_{i}\sigma_x^{(i)}+\frac{\Delta}{2}\sum_{i}\sigma_z^{(i)}+\frac{V}{4}\sum_{ij}\sigma_z^{(i)}\sigma_z^{(j)}
\end{align}
where $\Omega$ and $\Delta$ follow from the laser parameters and $V$
is the interaction strength between neighboring sites. In the following, we will focus on the situation where the
spin Hamiltonian describes an Ising model in a purely transverse
field, i.e., $\Delta=0$.

\begin{figure}[t!]
\centering
\includegraphics[width=1.0\linewidth]{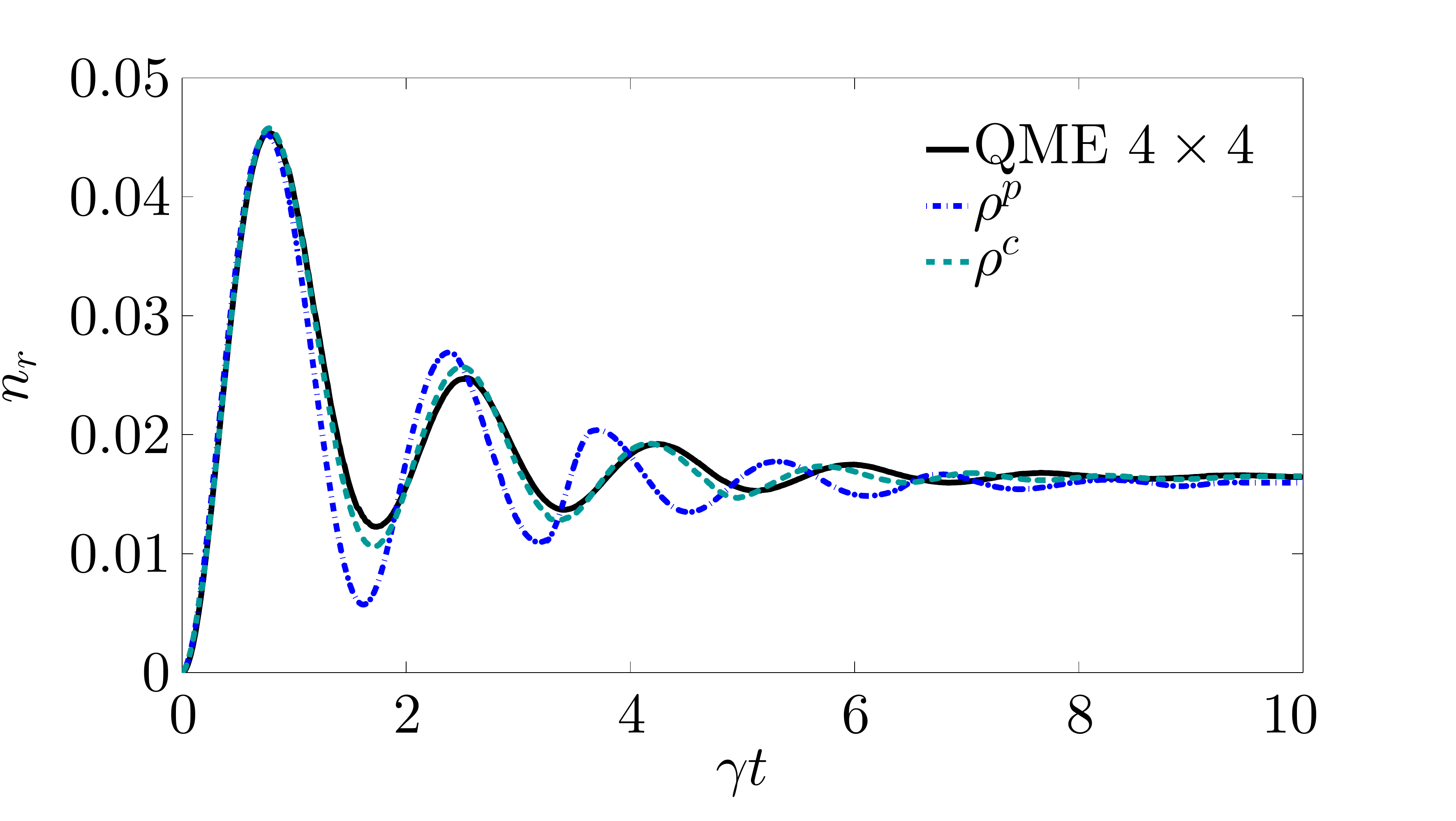}
\caption{(Color online) Time evolution of the Rydberg density $n_r$ calculated via
  the quantum trajectory method in a $4 \times 4$ lattice (solid) and
  the variational methods including correlations $\rho^c$ (dashed) and
  product states $\rho^p$ (dotted) ($\Omega=\gamma$, $V=2\,\gamma$).}
\label{W1V2}
\end{figure}
The initial state is fully polarized into the electronic ground state,
corresponding to the typical experimental situation. In the following,
we compare the variational solutions based on product states and on
states including nearest-neighbor correlations to a
quantum trajectory solution of the quantum master equation for a
$4\times 4$ lattice \cite{Johansson2013}. The resulting time evolution
of the average Rydberg density $n_r$ is shown in Fig.~1. While the
steady state value is consistent in all three approaches, the
variational product state solution shows significant deviations during
the relaxation dynamics. Remarkably, the variational approach based on
correlated states shows much better quantitative agreement with the full
numerical simulation for all times. This agreement is also found over
a large region of both $\Omega$ and $V$; see the Appendix.
  Only in the regime with $\Omega \gg \gamma$ and $V
\gg \gamma$ are there significant differences as the system undergoes
a liquid-gas phase transition of the steady state
\cite{Weimer2015}. However, in this regime, the quantum trajectories
solution cannot be expected to provide an accurate description of the
system in the thermodynamic limit, as strong finite size effects are
present in a $4 \times 4$ system \cite{Weimer2015a}. These results
demonstrate that our variational approach can be used to accurately
approximate the time evolution of open quantum many-body systems.
\section{Non-Markovianity and quantum linear mutual information}
Using our variational method, we can now investigate the dynamics of
the system in more detail. As the variational state involving two-site
correlations provides an accurate description, we focus on the
dynamics of the reduced two-site density operator $\rho_{ij}$ as our
subsystem of interest. The interaction with the environment formed by
the other sites results in additional coherent and dissipative
terms. In the most general case, the equation of motion for the
reduced density operator can be written in the form \cite{Hall2014}
\begin{align}
  \frac{d}{dt}\rho_{ij} & =-i[H(t),\rho_{ij}] \\ & \notag + \sum_{k=1}^{d^2-1} \gamma_k(t) \left(L_k^{\phantom \dagger}(t) \rho_{ij} L_k^{ \dagger}(t)-\frac{1}{2}\{L_k^{\phantom \dagger}(t) L_k^{\dagger}(t),\rho_{ij}\} \right)
\end{align}
with the dimension $d=4$ of the state space and the $L_k(t)$ forming an orthonormal basis according to
\begin{align}
 \mathrm{Tr}[L_k(t)]=0,~~~~\mathrm{Tr}[L_j^{\dagger}(t)L_k^{\phantom \dagger}(t)]=\delta_{jk},
\end{align}
and with the Hermitian operator $H(t)$ acting as an effective
Hamiltonian.  In contrast to the Liouvillian in Lindblad form of
Eq.~(1), the generalized jump operators $L_k(t)$ can now become
time-dependent, and the generalized decay rates $\gamma_k(t)$ may
become negative. The appearance of a negative decay rate corresponds
to the dynamics being non-Markovian; conversely, if all decoherence
rates are positive, the dynamics is Markovian. This relationship
provides a natural measure for the degree of non-Markovianity of the
dynamics,
\begin{align}
 f(t)=\frac{1}{2}\sum_{k=1}^{d^2-1}[|\gamma_k(t)|-\gamma_k(t)],
 \label{nonmarkov}
\end{align}
which is the sum of all negative decay rates \cite{Hall2014}. This
quantity is also closely related to other measures of non-Markovianity
\cite{Rivas2014}.

\begin{figure}[t!]
\centering
\includegraphics[width=\linewidth]{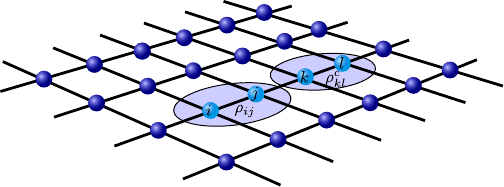}
\caption{(Color online) Lattice structure for the calculation of the non-Markovianity
  $f(t)$. The variational state $\rho_{ij}$ is being iterated over a
  complete set of initial states. The environment interacting with
  the sites $i$ and $j$ is given by the solution to the time evolution
  of the system for correlated states $\rho^c_{kl}$; see Fig.~\ref{W1V2}.}
\label{lattice}
\end{figure}

For the calculation of the generalized decay rates $\gamma_k(t)$, we
consider the dynamics of a complete set of linearly independent states,
\begin{align}
 \rho^{00}=\frac{\sigma_0 \otimes \sigma_0}{4};~~~   \rho^{mn}=\frac{1+\sigma_m \otimes \sigma_n}{4} \\ \notag m,n \in \{0,x,y,z\}; ~~ m+n \ne 0,
\end{align}
with $\sigma_0$ being the identity.
 While the reduced density operator
$\rho_{ij}$ is iterated over a complete set, the state of the
environment is fixed to the variational solution $\rho^c(t)$; see
Fig.~\ref{lattice}. This results in a consistent treatment of the
non-Markovianity of the dynamics, which is governed by the
correlations between the two-site system of interest and the
environment formed by the remaining sites.

\begin{figure}[t!]
\centering
\includegraphics[width=1.0\linewidth]{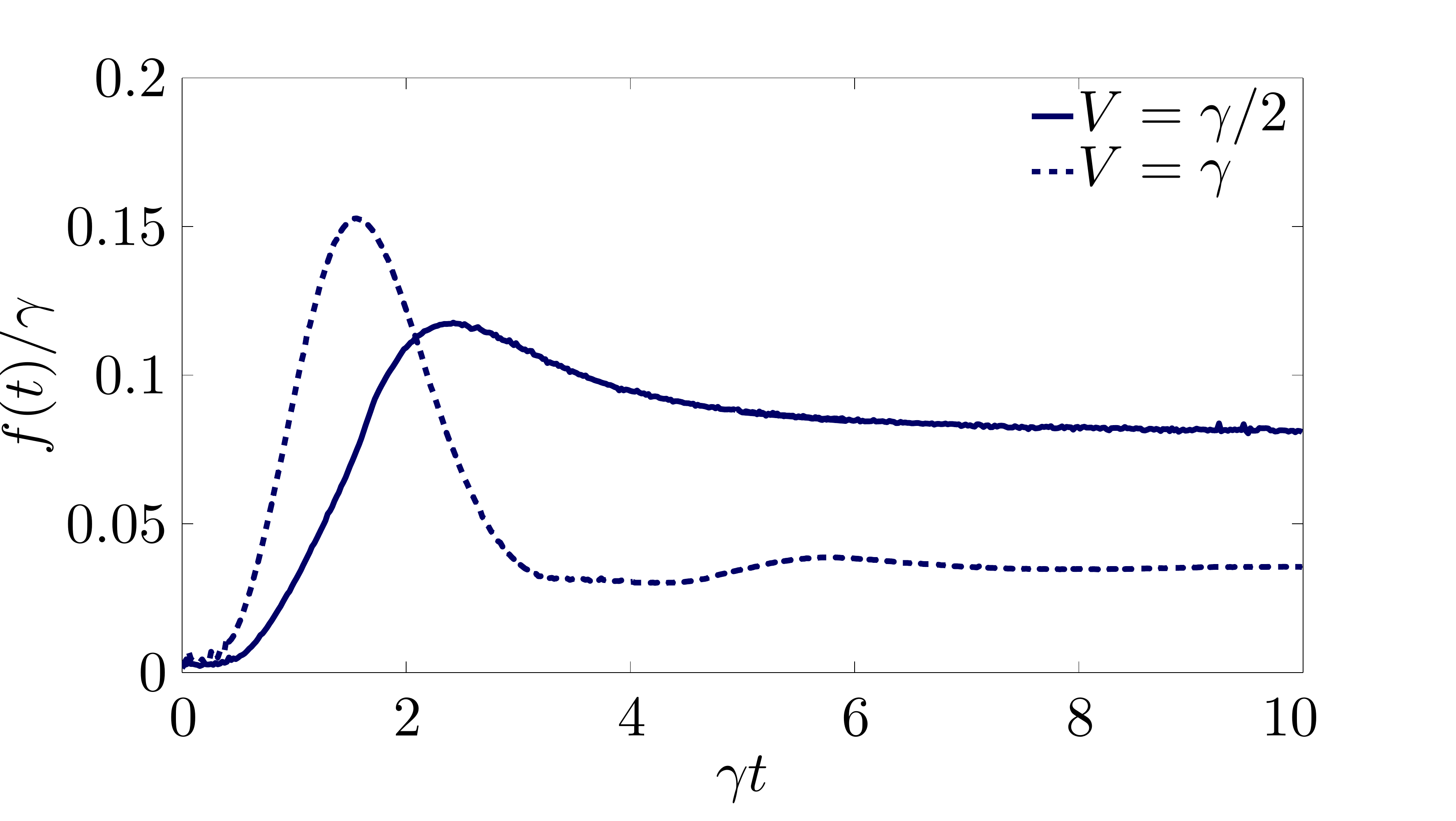}
\caption{(Color online) Time evolution of the non-Markovianity $f(t)$ for an
  interaction strength of $V= \gamma/2$ (solid) and $V=\gamma$
  (dashed) ($\Omega = \gamma$).}
\label{fig:markov}
\end{figure}

For the calculation of the non-Markovianity $f(t)$, we need to obtain
both $\rho_{ij}(t)$ and $\dot{\rho}_{ij}(t)$. Here, we can readily
compute $\rho_{ij}(t)$ using the variational method, from which we can
determine its derivative according to
\begin{align}
 \dot \rho_{ij}(t+\tau)=\frac{\rho_{ij}(t+2\tau)-\rho_{ij}(t)}{2\tau}+ O(\tau^3).
\end{align}
The dynamics can be rewritten in the form
\begin{align}
 \dot{\rho}_{ij}(t+\tau)=\sum_{\alpha \beta}c_{\alpha \beta}(t+\tau) G_{\alpha} \rho_{i j}(t+\tau) G_{\beta},
 \label{lineq}
\end{align}
which is also valid in the case of non-Markovian dynamics
\cite{Gorini1976}. Here, the $G_{\alpha}$ and $G_{\beta}$ form a
  complete set of Hermitian operators, i.e., all possible combinations
  of the tensor product of two Pauli matrices.  By iterating over all
possible initial states for $\rho_{ij}$, we can uniquely determine the
matrix elements $c_{\alpha\beta}(t)$, from which we can finally
calculate the generalized decay rates $\gamma_k(t)$
\cite{Hall2014}. Hence, the generalized decay
  rates and the effective jump operators $L_k(t)$ are calculated via
  the variational solution of the full quantum many-body model and
  differ from the purely Markovian jump operators in Eq.~(1). Then,
according to Eq.~(\ref{nonmarkov}), we find that the non-Markovianity
$f(t)$ is nonzero during the entire evolution; see
Fig.~\ref{fig:markov}. We also see that the degree of non-Markovianity
is the largest for intermediate times, which explains why product
states---which cannot generate non-Markovian dynamics---do not give an
accurate description of the relaxation dynamics. Remarkably, the
non-Markovianity even persists in the steady state.

Experimentally measuring the quantity $f(t)$ is a challenging task,
requiring us to separate the dynamics of the two sites of interest from
their environment. Therefore, we aim to construct a much more
accessible quantity, which can serve as a witness for the
non-Markovianity of the non-equilibrium steady state of the
system. Here, we find that the quantum linear mutual information
(QLMI), which depends only on the two-site density operator
$\rho_{ij}$ fulfills this property. The QLMI is a measure for the
quantum and classical correlations in the system and is defined as
\cite{Angelo2004}
\begin{align}
 I=S_l(\rho_i \otimes \rho_j)-S_l(\rho_{ij}),
\end{align}
according to the linear entropy
\begin{align}
 S_l(\rho_{ij})=1-\text{Tr}\{\rho_{ij}^2\}.
\end{align}

\begin{figure}[t!]
\centering
\includegraphics[width=1.0\linewidth]{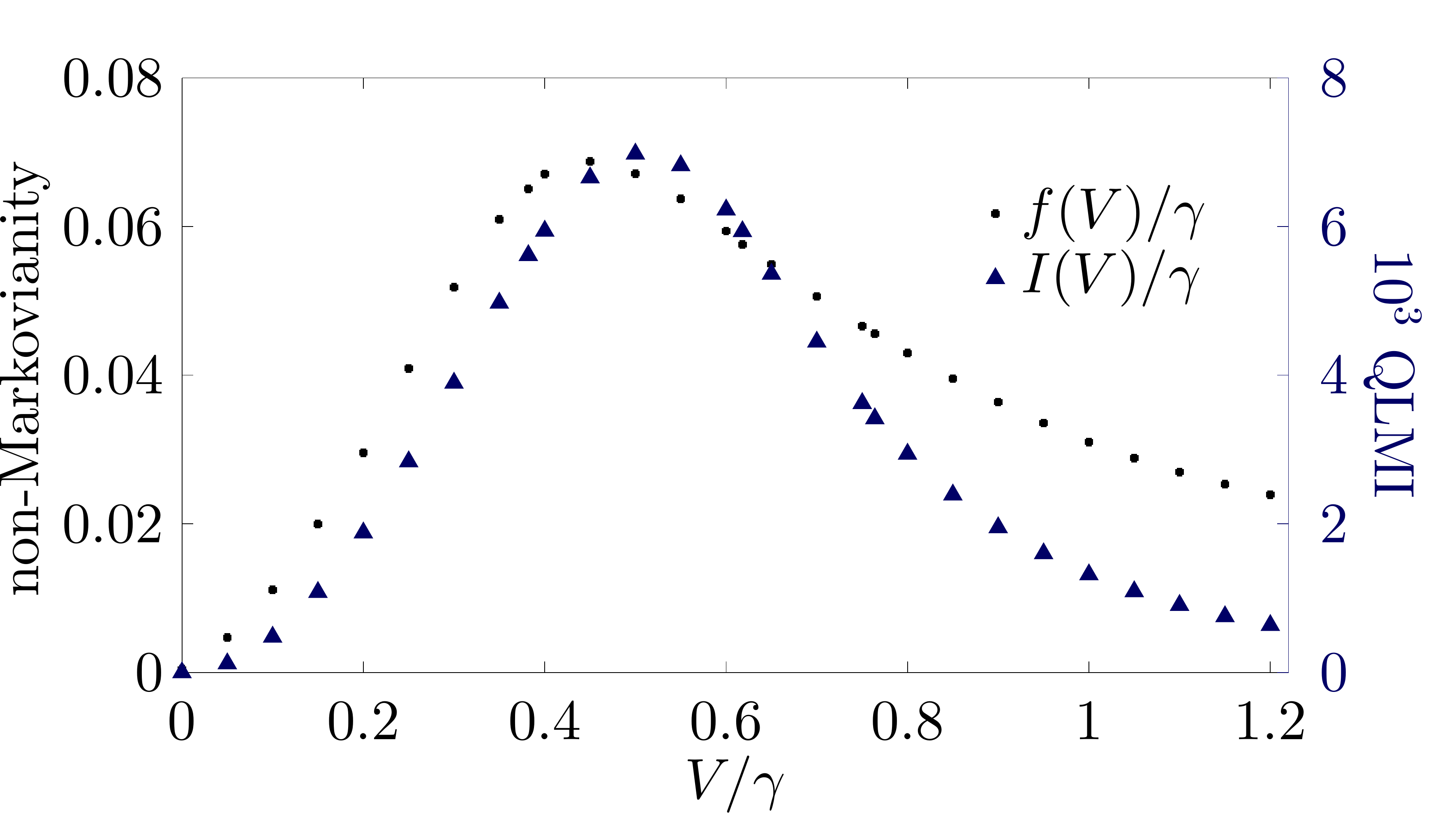}
\caption{(Color online) Non-Markovianity $f(V)$ and the QLMI $I(V)$ of the steady state for different values of the interaction strength $V$ ($\Omega=\gamma$).}
\label{fig:entropy}
\end{figure}

Consequently, the QLMI is a natural extension of the (linear)
entanglement entropy for mixed quantum states. As the QLMI is a
functional of the reduced density operator $\rho_{ij}$, it can be
experimentally determined using standard quantum state tomography. In
our case, the relationship between the QLMI and non-Markovianity can
be understood as both being intrinsically related to the correlations
present in the system. Indeed, we observe a good quantitative
agreement, up to a trivial constant factor, between the two measures,
see Fig.~\ref{fig:entropy}.

Remarkably, this quantitative agreement is only found for the QLMI; it
is absent for other measures such as the von Neumann mutual
information $I_{\mathrm{VN}}$, where the linear entropy $S_l$ is
replaced by the von Neumann entropy
$S(\rho)=-\text{Tr}\{\rho\log\rho\}$. We can understand this property
by considering an expansion of $\rho_{ij}$ around product states,
\begin{align}
 \rho_{ij}=\rho_i\otimes \rho_j+\varepsilon A
\end{align}
where $\varepsilon$ is the expansion parameter and $A =
\sigma_\kappa^{(i)} \sigma_\lambda^{(j)}$ is a tensor product of two Pauli
matrices. We now expand $\rho_i$ and $\rho_j$ in terms of the
variational parameters $\alpha_\mu$, according to
Eq.~(\ref{varstate}). Then, we find for the von Neumann mutual
information
\begin{align}
  I_{\mathrm{VN}} =  \varepsilon\,\text{Tr}\{A \ln (\rho_i \otimes \rho_j)\} +O(\varepsilon^2),
\end{align}
which leads to a logarithmic dependence on the variational parameters
$\alpha_\mu$. On the other hand, the non-Markovianity follows from
minimizing the variational norm $D$. Here, we find that both $D$ and
the variational solution for $\rho_{ij}(t+\tau)$ are a bilinear function
in terms of the parameters $\alpha_\mu$, which is incompatible with
the logarithmic dependence predicted by the von Neumann mutual
information. In contrast, the leading term of the QLMI remarkably
reproduces this bilinear form, i.e.,
\begin{equation}
  I=\varepsilon \alpha_\kappa \alpha_\lambda + O(\varepsilon^2).
\end{equation}
This singles out the QLMI as the correct information-theoretical
measure to serve as a witness for non-Markovianity in our analysis. We
also note that our focus on the linear contribution in $\varepsilon$
becomes less accurate for larger values of $V$: In the case of
sufficiently strong interactions, the coherent part of the dynamics
essentially becomes frozen and the stationary state lies close to the
pure state with all atoms being polarized into their electronic ground
state. This is accompanied by a vanishing of the first order
contribution to the QLMI, as all coefficients except for $\alpha_z$
approach zero. In this regime, the QLMI and the non-Markovianity
indeed begin to deviate from each other; see Fig.~\ref{fig:entropy},
with the QLMI decaying faster than the non-Markovianity.

\section{Summary}
In summary, we have successfully demonstrated a variational
method for calculating the time evolution of open quantum many-body
systems. Our method allows for a systematic treatment of correlations
in the system, and thus gives access to evaluate quantities associated
with non-Markovian dynamics. Finally, we wish to point out that our
variational method is also applicable to the time evolution of closed
quantum many-body systems in the absence of dissipation. Crucially,
the description in terms of correlated variational states generically
leads to an effective equation of motion that contains dissipative
terms that are formed by tracing out the environment formed by the
other sites of the system. We thus expect our method to be highly
relevant for the investigation of quench dynamics
\cite{Kollath2007,Manmana2007,Trotzky2012}, thermalization in closed
quantum systems \cite{Rigol2008,Langen2015}, or quantum control of
many-body systems \cite{Doria2011}.

\begin{acknowledgments}

  We acknowledge fruitful discussions with O.~Morsch and
  T.~Osborne. This work was funded by the Volkswagen Foundation and
  the DFG (Research Training Group 1729).

\end{acknowledgments}

\onecolumngrid

 \begin{figure}[H]
 \includegraphics[width=0.993\linewidth]{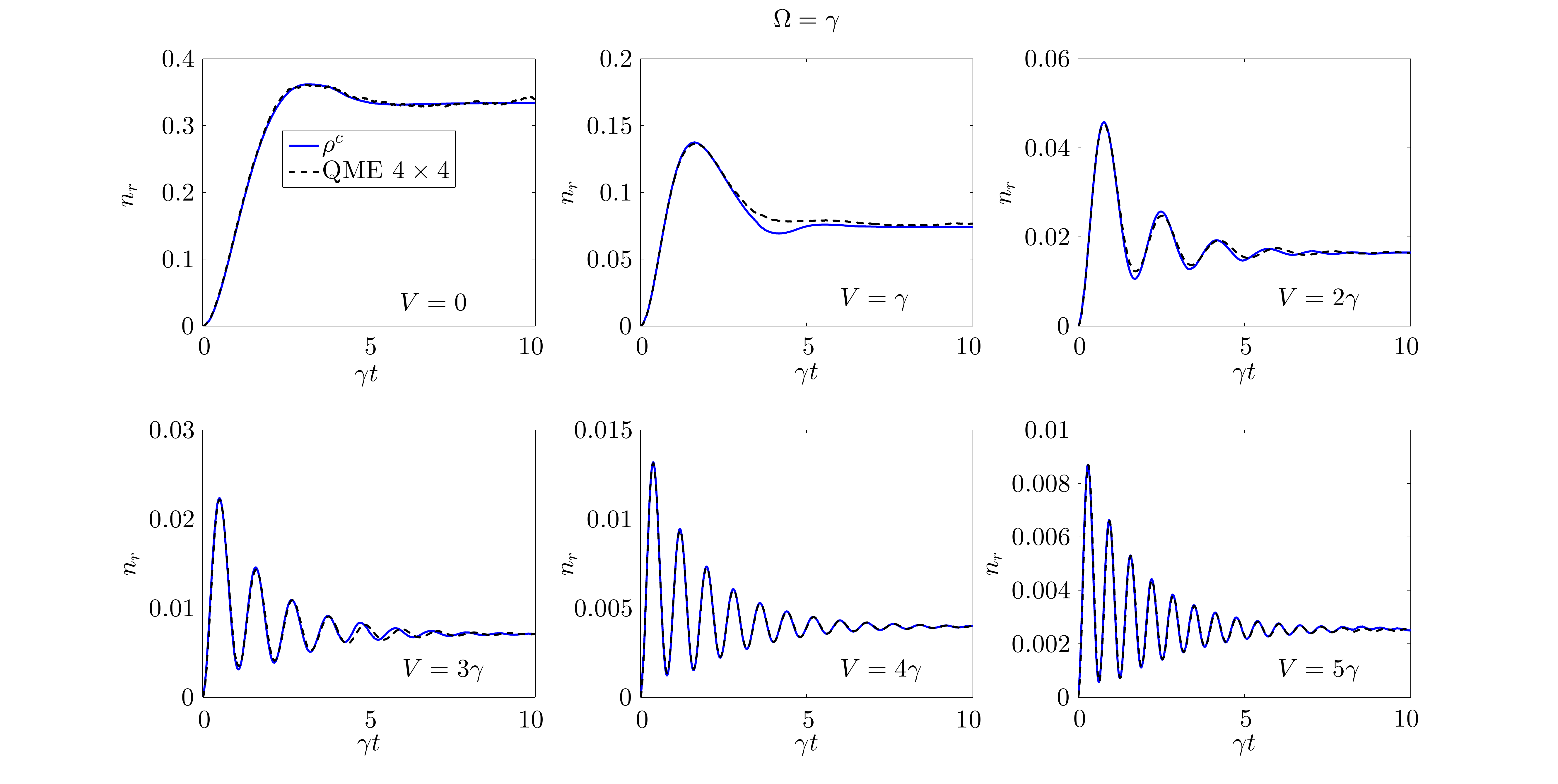}
 \caption{(Color online) Dynamics of the Rydberg density $n_r$ for $\Omega=\gamma$,
   calculated by the variational approach (solid) and the quantum
   trajectory method (dashed).}
 \label{W=1}
\end{figure}

\begin{figure}[H]
 \includegraphics[width=0.993\linewidth]{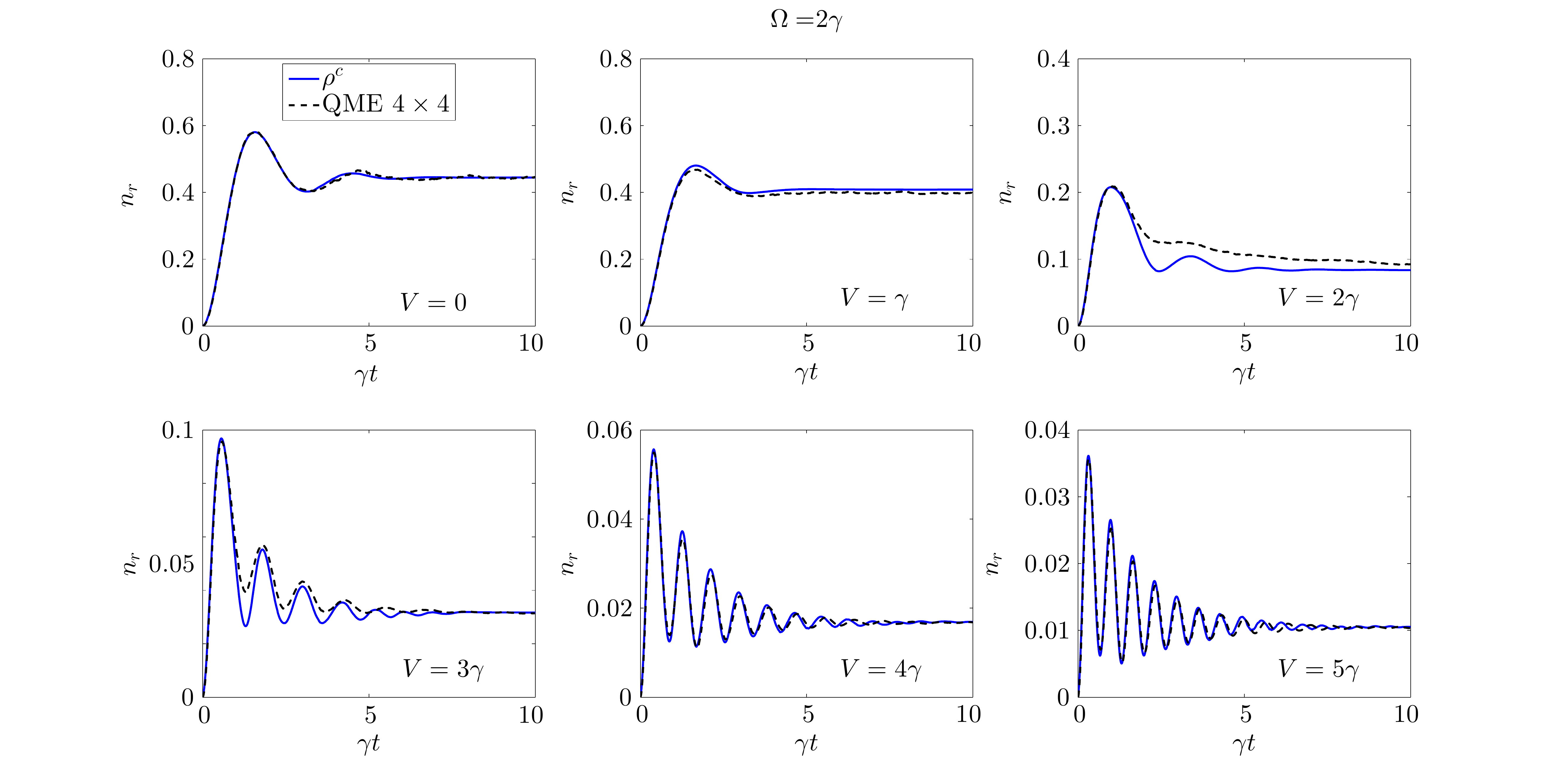}
 \caption{(Color online) Rydberg density $n_r$ for $\Omega=2\,\gamma$ and several values of the interaction strength $V$.}
 \label{W=2}
\end{figure}

\twocolumngrid

\appendix
\setcounter{secnumdepth}{0}
\section{APPENDIX: Comparison of the variational method and the quantum trajectory method}

\begin{figure}[H]
 \includegraphics[width=1\linewidth]{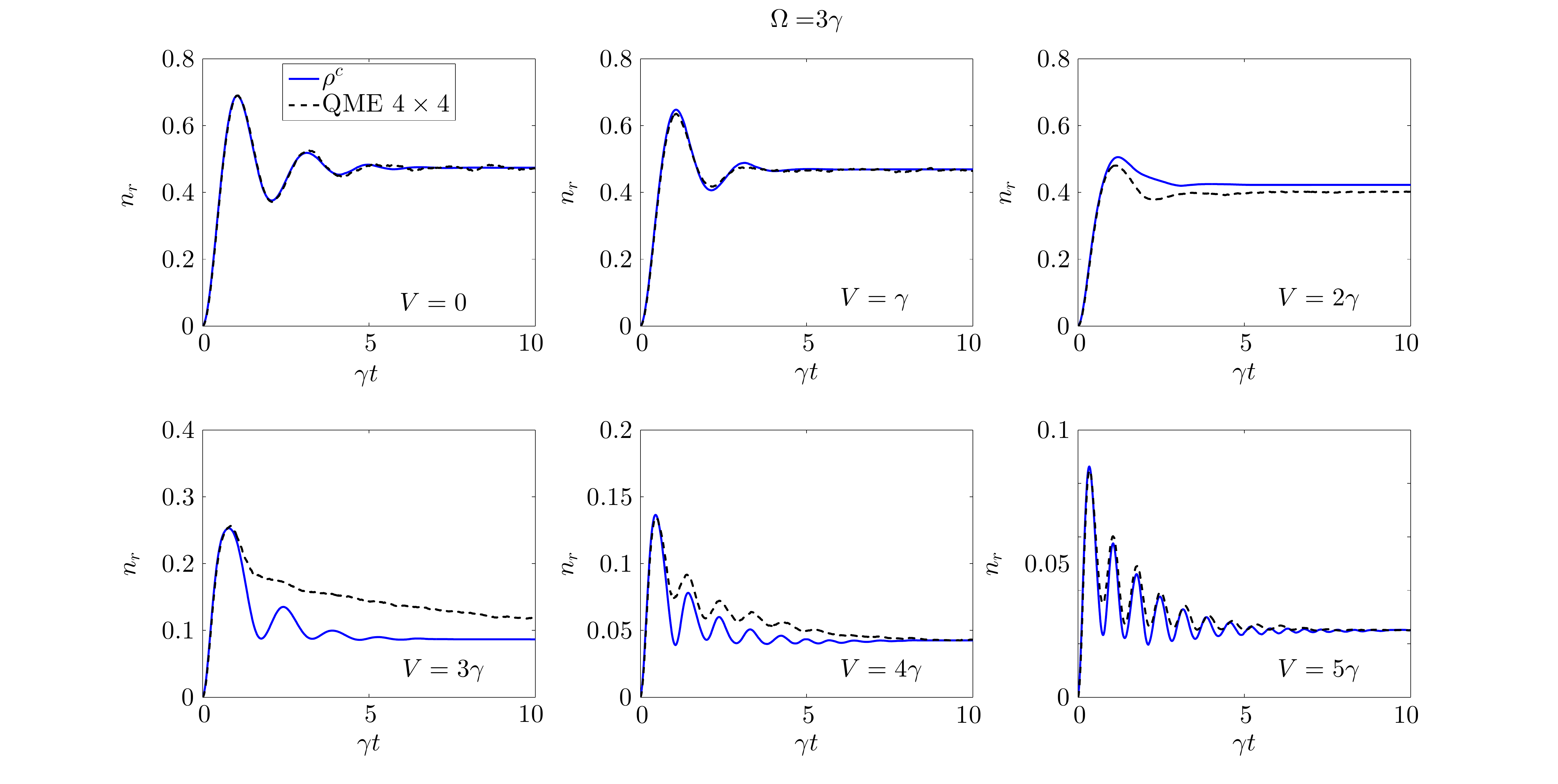}
 \caption{(Color online) Rydberg density for $\Omega=3\,\gamma$ and several values of $V$.}
 \label{W=3}
\end{figure}

\begin{figure}[H]
 \includegraphics[width=1\linewidth]{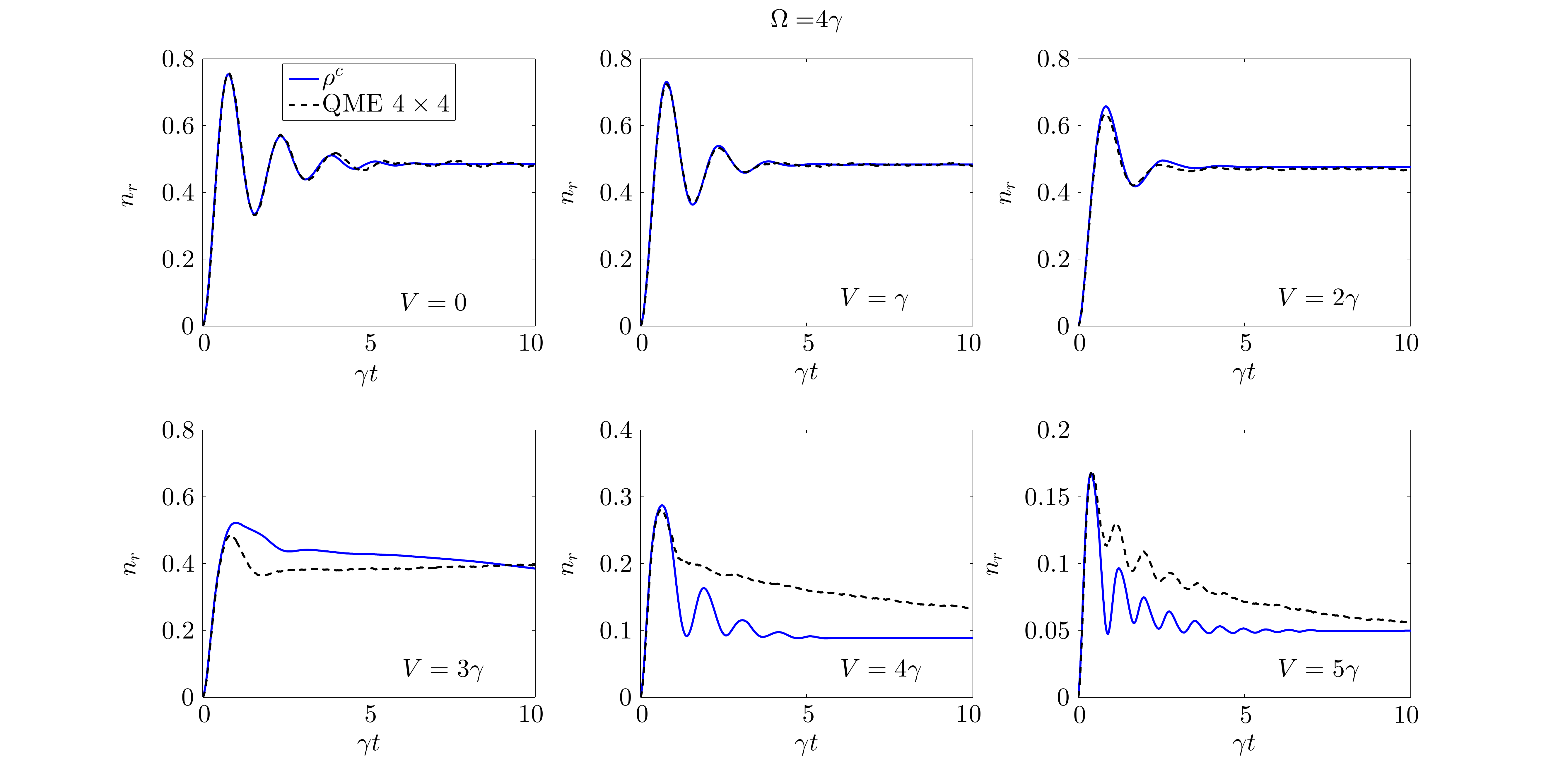}
 \caption{(Color online) Rydberg density for $\Omega=4\,\gamma$ and several values of $V$.}
 \label{W=4}
\end{figure}

In the following, we compare the dynamics gained via the variational
principle with the dynamics of the full quantum master equation. The
latter is calculated by the quantum trajectory method on a
two-dimensional $4 \times 4$ lattice with periodic boundary
conditions. Figs.~\ref{W=1} to \ref{W=4} show the dynamics of the Rydberg density
for $\Omega=\gamma$ to $\Omega=4\,\gamma$, respectively, for different
interaction strengths $V$.

\onecolumngrid

\twocolumngrid

  For larger values of $\Omega$ and $V$, the
system undergoes a phase transition of the nonequilibrium steady state
\cite{Weimer2015}. 
In that region, the variational solution does not
match very well with the solution of the full quantum master equation
anymore, as can be seen for the case of $\Omega=3\,\gamma$ and
$\Omega=4\,\gamma$. 
This can be traced back to two reasons: (i) Close
to the phase transitions, long-range correlations become relevant,
whereas our variational approach only takes nearest-neighbour
correlations into account. (ii) At the same time, the quantum trajectories
solution is subject to finite size effects, which also become
important close to the transition.



\begin{thebibliography}{10}


\bibitem{Breuer2002}
H.-P. Breuer and F.~Petruccione,
\newblock {\em The {t}heory of {o}pen {q}uantum {s}ystems} (Oxford University
  Press, Oxford, 2002).

\bibitem{Syassen2008}
N.~Syassen, D.~M. Bauer, M.~Lettner, T.~Volz, D.~Dietze, J.~J. Garc\'ia-Ripoll,
  J.~I. Cirac, G.~Rempe, and S.~D\"urr,
\newblock Strong dissipation inhibits losses and induces correlations in cold
  molecular gases,
\newblock Science {\bf 320}, 1329 (2008).

\bibitem{Baumann2010}
K.~{Baumann}, C.~{Guerlin}, F.~{Brennecke}, and T.~{Esslinger},
\newblock {Dicke quantum phase transition with a superfluid gas in an optical
  cavity},
\newblock Nature (London) {\bf 464}, 1301 (2010).

\bibitem{Barreiro2011}
J.~T. Barreiro, M.~M\"uller, P.~Schindler, D.~Nigg, T.~Monz, M.~Chwalla,
  M.~Hennrich, C.~F. Roos, P.~Zoller, and R.~Blatt,
\newblock An open-system quantum simulator with trapped ions,
\newblock Nature (London) {\bf 470}, 486 (2011).

\bibitem{Krauter2011}
H.~Krauter, C.~A. Muschik, K.~Jensen, W.~Wasilewski, J.~M. Petersen, J.~I.
  Cirac, and E.~S. Polzik,
\newblock Entanglement Generated by Dissipation and Steady State Entanglement
  of Two Macroscopic Objects,
\newblock Phys. Rev. Lett. {\bf 107}, 080503 (2011).

\bibitem{Barontini2013}
G.~Barontini, R.~Labouvie, F.~Stubenrauch, A.~Vogler, V.~Guarrera, and H.~Ott,
\newblock Controlling the Dynamics of an Open Many-Body Quantum System with
  Localized Dissipation,
\newblock Phys. Rev. Lett. {\bf 110}, 035302 (2013).

\bibitem{Diehl2008}
S.~Diehl, {A. Micheli}, {A. Kantian}, {B. Kraus}, {H. P. B\"uchler}, and {P.
  Zoller},
\newblock Quantum states and phases in driven open quantum systems with cold
  atoms,
\newblock Nat. Phys. {\bf 4}, 878 (2008).

\bibitem{Verstraete2009}
F.~Verstraete, M.~M. Wolf, and J.~Ignacio~Cirac,
\newblock Quantum computation and quantum-state engineering driven by
  dissipation,
\newblock Nat. Phys. {\bf 5}, 633 (2009).

\bibitem{Weimer2010}
H.~{Weimer}, M.~{M{\"u}ller}, I.~{Lesanovsky}, P.~{Zoller}, and H.~P.
  {B{\"u}chler},
\newblock A Rydberg quantum simulator,
\newblock Nat. Phys. {\bf 6}, 382 (2010).

\bibitem{Raitzsch2009}
U.~Raitzsch, R.~Heidemann, H.~Weimer, B.~Butscher, P.~Kollmann, R.~L\"ow, H.~P.
  B\"uchler, and T.~Pfau,
\newblock Investigation of dephasing rates in an interacting Rydberg gas,
\newblock New J. Phys. {\bf 11}, 055014 (2009).

\bibitem{Carr2013}
C.~Carr, R.~Ritter, C.~G. Wade, C.~S. Adams, and K.~J. Weatherill,
\newblock Nonequilibrium Phase Transition in a Dilute Rydberg Ensemble,
\newblock Phys. Rev. Lett. {\bf 111}, 113901 (2013).

\bibitem{Malossi2014}
N.~Malossi, M.~M. Valado, S.~Scotto, P.~Huillery, P.~Pillet, D.~Ciampini,
  E.~Arimondo, and O.~Morsch,
\newblock Full Counting Statistics and Phase Diagram of a Dissipative Rydberg
  Gas,
\newblock Phys. Rev. Lett. {\bf 113}, 023006 (2014).

\bibitem{Schempp2014}
H.~Schempp, G.~G\"unter, M.~Robert-de Saint-Vincent, C.~S. Hofmann, D.~Breyel,
  A.~Komnik, D.~W. Sch\"onleber, M.~G\"arttner, J.~Evers, S.~Whitlock, and
  M.~Weidem\"uller,
\newblock Full Counting Statistics of Laser Excited Rydberg Aggregates in a
  One-Dimensional Geometry,
\newblock Phys. Rev. Lett. {\bf 112}, 013002 (2014).

\bibitem{Urvoy2015}
A.~Urvoy, F.~Ripka, I.~Lesanovsky, D.~Booth, J.~P. Shaffer, T.~Pfau, and
  R.~L\"ow,
\newblock Strongly Correlated Growth of Rydberg Aggregates in a Vapor Cell,
\newblock Phys. Rev. Lett. {\bf 114}, 203002 (2015).

\bibitem{Weber2015}
T.~M. Weber, M.~Honing, T.~Niederprum, T.~Manthey, O.~Thomas, V.~Guarrera,
  M.~Fleischhauer, G.~Barontini, and H.~Ott,
\newblock {Mesoscopic} {Rydberg-blockaded} ensembles in the superatom regime
  and beyond,
\newblock Nat. Phys. {\bf 11}, 157 (2015).

\bibitem{Low2012}
R.~L\"ow, H.~Weimer, J.~Nipper, J.~B. Balewski, B.~Butscher, H.~P. B\"uchler,
  and T.~Pfau,
\newblock An experimental and theoretical guide to strongly interacting Rydberg
  gases,
\newblock J. Phys. B {\bf 45}, 113001 (2012).

\bibitem{Lee2011}
T.~E. Lee, H.~H\"affner, and M.~C. Cross,
\newblock Antiferromagnetic phase transition in a nonequilibrium lattice of
  Rydberg atoms,
\newblock Phys. Rev. A {\bf 84}, 031402 (2011).

\bibitem{Honer2011}
J.~Honer, R.~L\"ow, H.~Weimer, T.~Pfau, and H.~P. B\"uchler,
\newblock Artificial Atoms Can Do More Than Atoms: Deterministic Single Photon
  Subtraction from Arbitrary Light Fields,
\newblock Phys. Rev. Lett. {\bf 107}, 093601 (2011).

\bibitem{Glatzle2012}
A.~W. Glaetzle, R.~Nath, B.~Zhao, G.~Pupillo, and P.~Zoller,
\newblock Driven-dissipative dynamics of a strongly interacting Rydberg gas,
\newblock Phys. Rev. A {\bf 86}, 043403 (2012).

\bibitem{Ates2012a}
C.~Ates, B.~Olmos, J.~P. Garrahan, and I.~Lesanovsky,
\newblock Dynamical phases and intermittency of the dissipative quantum Ising
  model,
\newblock Phys. Rev. A {\bf 85}, 043620 (2012).

\bibitem{Lemeshko2013a}
M.~{Lemeshko} and H.~{Weimer},
\newblock {Dissipative binding of atoms by non-conservative forces},
\newblock Nat. Commun. {\bf 4}, 2230 (2013).

\bibitem{Hu2013}
A.~Hu, T.~E. Lee, and C.~W. Clark,
\newblock Spatial correlations of one-dimensional driven-dissipative systems of
  Rydberg atoms,
\newblock Phys. Rev. A {\bf 88}, 053627 (2013).

\bibitem{Honing2013}
M.~H\"oning, D.~Muth, D.~Petrosyan, and M.~Fleischhauer,
\newblock Steady-state crystallization of Rydberg excitations in an optically
  driven lattice gas,
\newblock Phys. Rev. A {\bf 87}, 023401 (2013).

\bibitem{Otterbach2014}
J.~Otterbach and M.~Lemeshko,
\newblock Dissipative Preparation of Spatial Order in Rydberg-Dressed
  Bose-Einstein Condensates,
\newblock Phys. Rev. Lett. {\bf 113}, 070401 (2014).

\bibitem{Sanders2014}
J.~Sanders, R.~van Bijnen, E.~Vredenbregt, and S.~Kokkelmans,
\newblock Wireless Network Control of Interacting Rydberg Atoms,
\newblock Phys. Rev. Lett. {\bf 112}, 163001 (2014).

\bibitem{Hoening2014}
M.~Hoening, W.~Abdussalam, M.~Fleischhauer, and T.~Pohl,
\newblock Antiferromagnetic long-range order in dissipative Rydberg lattices,
\newblock Phys. Rev. A {\bf 90}, 021603 (2014).

\bibitem{Marcuzzi2014}
M.~Marcuzzi, E.~Levi, S.~Diehl, J.~P. Garrahan, and I.~Lesanovsky,
\newblock Universal Nonequilibrium Properties of Dissipative Rydberg Gases,
\newblock Phys. Rev. Lett. {\bf 113}, 210401 (2014).

\bibitem{Weimer2015}
H.~Weimer,
\newblock Variational Principle for Steady States of Dissipative Quantum
  Many-Body Systems,
\newblock Phys. Rev. Lett. {\bf 114}, 040402 (2015).

\bibitem{Suli2003}
E.~S{\"u}li and D.~F. Mayers,
\newblock {\em An introduction to numerical analysis} (Cambridge university
  press, Cambridge, 2003).

\bibitem{Transchel2014}
F.~W.~G. {Transchel}, A.~{Milsted}, and T.~J. {Osborne},
\newblock {A monte carlo time-dependent variational principle},
\newblock {a}rXiv:1411.5546  [quant-ph].

\bibitem{Cui2015}
J.~Cui, J.~I. Cirac, and M.~C. Ba\~nuls,
\newblock Variational Matrix Product Operators for the Steady State of
  Dissipative Quantum Systems,
\newblock Phys. Rev. Lett. {\bf 114}, 220601 (2015).

\bibitem{Mascarenhas2015}
E.~Mascarenhas, H.~Flayac, and V.~Savona,
\newblock Matrix-product-operator approach to the nonequilibrium steady state of driven-dissipative quantum arrays,
\newblock Phys. Rev. A {\bf 92}, 022116 (2015).

\bibitem{Tomadin2010}
A.~Tomadin, V.~Giovannetti, R.~Fazio, D.~Gerace, I.~Carusotto, H.~E. T\"ureci,
  and A.~Imamoglu,
\newblock Signatures of the superfluid-insulator phase transition in
  laser-driven dissipative nonlinear cavity arrays,
\newblock Phys. Rev. A {\bf 81}, 061801 (2010).

\bibitem{Poletti2013}
D.~Poletti, P.~Barmettler, A.~Georges, and C.~Kollath,
\newblock Emergence of Glasslike Dynamics for Dissipative and Strongly
  Interacting Bosons,
\newblock Phys. Rev. Lett. {\bf 111}, 195301 (2013).

\bibitem{Vidanovic2014}
I.~Vidanovi\ifmmode~\acute{c}\else \'{c}\fi{}, D.~Cocks, and W.~Hofstetter,
\newblock Dissipation through localized loss in bosonic systems with long-range
  interactions,
\newblock Phys. Rev. A {\bf 89}, 053614 (2014).

\bibitem{Weimer2015a}
H.~Weimer,
\newblock Variational analysis of driven-dissipative Rydberg gases,
\newblock Phys. Rev. A {\bf 91}, 063401 (2015).

\bibitem{Maghrebi2015}
M.~F. Maghrebi, and A.~V. Gorshkov,
\newblock Nonequilibrium many-body steady states via Keldysh formalism,
\newblock {a}rXiv:1507.01939 [cond-mat.quant-gas].

\bibitem{Mendoza-Arenas2015}
J.~J. Mendoza-Arenas, S.~R. Clark, S.~Felicetti, G.~Romero, and E.~Solano, D.~G. Angelakis, and D.~Jaksch,
\newblock Beyond mean-field bistability in driven-dissipative lattices: bunching-antibunching transition and quantum simulation,
\newblock {a}rXiv:1510.06651 [quant-ph].

\bibitem{Johansson2013}
J.~Johansson, P.~Nation, and F.~Nori,
\newblock QuTiP 2: A Python framework for the dynamics of open quantum systems,
\newblock Comp. Phys. Comm. {\bf 184}, 1234 (2013).

\bibitem{Hall2014}
M.~J.~W. Hall, J.~D. Cresser, L.~Li, and E.~Andersson,
\newblock Canonical form of master equations and characterization of
  non-Markovianity,
\newblock Phys. Rev. A {\bf 89}, 042120 (2014).

\bibitem{Rivas2014}
A.~Rivas, S.~F. Huelga, and M.~B. Plenio,
\newblock Quantum non-Markovianity: characterization, quantification and
  detection,
\newblock Rep. Prog. Phys. {\bf 77}, 094001 (2014).

\bibitem{Gorini1976}
V.~Gorini, A.~Kossakowski, and E.~C.~G. Sudarshan,
\newblock Completely positive dynamical semigroups of N-level systems,
\newblock J. Math. Phys. {\bf 17}, 821 (1976).

\bibitem{Angelo2004}
R.~M. Angelo, S.~A. Vitiello, M.~A.~M. de~Aguiar, and K.~Furuya,
\newblock Quantum linear mutual information and classical correlations in
  globally pure bipartite systems,
\newblock Physica A {\bf 338}, 458 (2004).

\bibitem{Kollath2007}
C.~Kollath, A.~M. L\"auchli, and E.~Altman,
\newblock Quench Dynamics and Nonequilibrium Phase Diagram of the Bose-Hubbard
  Model,
\newblock Phys. Rev. Lett. {\bf 98}, 180601 (2007).

\bibitem{Manmana2007}
S.~R. Manmana, S.~Wessel, R.~M. Noack, and A.~Muramatsu,
\newblock Strongly Correlated Fermions after a Quantum Quench,
\newblock Phys. Rev. Lett. {\bf 98}, 210405 (2007).

\bibitem{Trotzky2012}
S.~Trotzky, Y.-A. Chen, A.~Flesch, I.~P. McCulloch, U.~Schollwock, J.~Eisert,
  and I.~Bloch,
\newblock Probing the relaxation towards equilibrium in an isolated strongly
  correlated one-dimensional {Bose} gas,
\newblock Nat. Phys. {\bf 8}, 325 (2012).

\bibitem{Rigol2008}
M.~Rigol, V.~Dunjko, and M.~Olshanii,
\newblock Thermalization and its mechanism for generic isolated quantum
  systems,
\newblock Nature (London) {\bf 452}, 854 (2008).

\bibitem{Langen2015}
T.~Langen, S.~Erne, R.~Geiger, B.~Rauer, T.~Schweigler, M.~Kuhnert,
  W.~Rohringer, I.~E. Mazets, T.~Gasenzer, and J.~Schmiedmayer,
\newblock Experimental observation of a generalized Gibbs ensemble,
\newblock Science {\bf 348}, 207 (2015).

\bibitem{Doria2011}
P.~Doria, T.~Calarco, and S.~Montangero,
\newblock Optimal Control Technique for Many-Body Quantum Dynamics,
\newblock Phys. Rev. Lett. {\bf 106}, 190501 (2011).

\end{thebibliography}
\end{document}